\newcommand{\beq}{\begin{equation}}
\newcommand{\eeq}{\end{equation}}

\documentclass{ws-mplb}

\begin{document}

\markboth{Kevrekidis and Frantzeskakis}{PATTERN FORMING DYNAMICAL INSTABILITIES OF BOSE-EINSTEIN CONDENSATES}

%
%

\title{PATTERN FORMING DYNAMICAL INSTABILITIES OF BOSE-EINSTEIN CONDENSATES: A SHORT REVIEW}

\author{P.G. KEVREKIDIS}

\address{Department of Mathematics and Statistics, 
University of Massachusetts, \\
Amherst, MA 01003, USA \\
kevrekid@math.umass.edu}

\author{D.J. FRANTZESKAKIS}

\address{Department of Physics, University of Athens, 
Panepistimiopolis, \\
Zografos, Athens 15784, Greece\\
dfrantz@cc.uoa.gr}

\maketitle

\begin{history}
\received{03/15/04}
\end{history}

\begin{abstract}
In this short topical review, we revisit a number of works on the 
pattern-forming 
dynamical instabilities of Bose-Einstein condensates  
in one- and two-dimensional settings. In particular, 
we illustrate the trapping conditions that allow the reduction of
the three-dimensional, mean field description of the condensates
(through the Gross-Pitaevskii equation) to such lower dimensional
settings, as well as to lattice settings. We then go on to study
the modulational instability in one dimension and the snaking/transverse
instability in two dimensions as typical examples of long-wavelength
perturbations that can destabilize the condensates and lead to the
formation of patterns of coherent structures in them. 
Trains of 
solitons in one-dimension and vortex arrays in two-dimensions are prototypical
examples of the resulting nonlinear waveforms, upon which we 
briefly touch at the end of this review.
\end{abstract}

\section{Abbreviations}

\begin{itemlist}
\item BEC: Bose-Einstein Condensate
\item FRM: Feshbach Resonance Management
\item GP: Gross-Pitaevskii (Equation)
\item MI: Modulational Instability
\item NLS: Nonlinear Schr{\"o}dinger (Equation)
\item OL: Optical lattice
\item RDS: Ring Dark Soliton
\item SI: Snaking Instability
\item TF: Thomas-Fermi
\item TI: Transverse Instability
\end{itemlist}

\section{Overview}

Bose-Einstein condensation (BEC) was initially predicted by Bose and Einstein in 1924 
\cite{bose,einstein1,einstein2}. It involves the macroscopic occupation of the ground state of a gas of bosons, below a critical transition temperature $T_c$, i.e., a quantum phase transition. It took, however, over 70 years for this prediction to be experimentally verified by a fascinating series of experiments in 1995 in vapors of rubidium \cite{anderson} and sodium \cite{davis}. In the same year, first signatures of the occurrence of BEC were also reported in vapors of lithium \cite{bradley} (and were later more systematically confirmed). The ability of the experimental groups to controllably cool alkali atoms (currently over $35$ groups around the world can routinely produce BECs) at sufficiently low temperatures and confine them via a combination of magnetic and optical techniques (for a review see e.g., \cite{review,book}), has been instrumental in this major feat whose significance has already been recognized through the 2001 Nobel prize in Physics.

This development is of particular interest also from a theoretical/mathematical standpoint. 
On the one hand, there is a detailed and ``clean'' experimental control over the produced Bose condensates. On the other hand, 
perhaps more importantly, for experimentally relevant conditions, there exists a very good model partial differential equation (PDE) that can describe, at the mean field level, the behavior of the condensate. This mean-field model (which, at heart, approximates a quantum many-body interaction with a classical, but nonlinear self-interaction 
\cite{lieb,lieb1}) is the well-known Gross-Pitaevskii (GP) equation \cite{gross,pittaevskii}. The latter is a variant of the famous Nonlinear Schr{\"o}dinger equation (NLS) \cite{sulem}, incorporating an external trapping potential, and reads:
\begin{equation}
i \hbar \Psi_t=-\frac{{\hbar}^2}{2 m} \Delta \Psi + g |\Psi|^2 \Psi + 
V_{{\rm ext}} ({\bf r}) \Psi,
\label{peq1}
\end{equation}
where $\Psi$ is the condensate wavefunction (the atom density is proportional to 
$|\Psi(x,t)|^2$), $\Delta$ is the Laplacian, $m$ is the atomic mass, 
and the prefactor $g$, which is proportional to the atomic scattering length \cite{review},  may take either positive (e.g., for rubidium and sodium) or negative (e.g., for lithium) values. The two cases correspond respectively to repulsive and attractive interactions between
the atoms, or to defocusing and focusing (respectively) Kerr nonlinearities in the language of nonlinear optics \cite{sulem}. Notice, however, that experimental ``wizardry'' can even manipulate the scattering length using the so-called Feshbach resonances \cite{feschb} to achieve any positive or negative value of the scattering length (i.e., nonlinearity strength in Eq.\ (\ref{peq1})). Moreover, the external potential can assume different forms. For the ``standard'' magnetic trap, this potential has a typical harmonic form:
\begin{equation}
V_{{\rm ext}}=\frac{m}{2} ( \omega_x^2 x^2 + \omega_y^2 y^2 + \omega_z^2 z^2),
\label{peq2}
\end{equation}
where, in general, the trap frequencies $\omega_x, \omega_y, \omega_z$ along the three directions are different. As a result, in recent experiments, the shape of the trap (and, hence, the form of the condensate itself) can range from isotropic forms to so-called cigar
shaped traps \cite{review}, to quasi two-dimensional ($2$D) \cite{catal1}, or even quasi 
one-dimensional ($1$D) \cite{catal2} forms. Moreover, linear ramps of (gravitational) potential 
$V_{{\rm ext}} = m g z$ have also been experimentally applied \cite{kasevich}. Another prominent example of an experimentally feasible (and controllable) potential is imposed by a pair of laser beams forming a standing wave which generates a periodic potential, the so-called optical lattice (OL) \cite{ol1,ol2,tromb,konot}, of the form:
\begin{equation}
V_{OL}=V_0 \left[\cos^2(\frac{2 \pi x}{\lambda_{x}} + \phi_x)
+ \cos^2(\frac{2 \pi y}{\lambda_{y}} + \phi_y)
+ \cos^2(\frac{2 \pi z}{\lambda_{z}} + \phi_z) \right],
\label{peq3}
\end{equation}
where $\lambda_{x,y,z}=\lambda \sin(\theta_{x,y,z}/2) /2$, $\lambda$ is the laser wavelength, $\theta$ is the (potentially variable) angle between the laser beams \cite{morsch} and 
$\phi_{x,y,z}$ 
are arbitrary phases. Such potentials have been realized in one 
\cite{ol1,ol2}, two (the so-called egg-carton potential) \cite{haensch} and three dimensions  \cite{catal1,greiner}. 

Moreover, present experimental realizations render feasible/controllable the adiabatic or abrupt displacement of the magnetic or optical lattice trap \cite{morsch,kas2} (inducing motion of the condensates), the phase engineering of the condensates in order to create vortices \cite{vort1,williams} or dark matter-wave solitons in them 
\cite{denschl,dark1,dark,dutton} 
the stirring (or rotation) of the condensates providing angular momentum and creating vortices 
\cite{vort2,vort3} and vortex-lattices \cite{latt1,latt2,latt3}, or the change of scattering length (from repulsive to attractive via Feshbach resonances) to produce bright matter-wave solitons and soliton trains \cite{bright1,bright2,njp2003b}. Both matter-wave solitons and vortices (which can been thought of as the $2$D analog of dark solitons) are fundamental nonlinear excitations of condensates, and as such have attracted much attention, as can be 
inferred from the relevant experimental and theoretical work. 
In addition, apart from being inherently 
interesting in the BEC physics, these nonlinear excitations are also relevant
to other fundamental issues of low-temperature (and soft condensed-matter) 
physics: for example we note the dominant role of vortices in the 
breakdown of superflow in Bose fluids \cite{jac11,jac12,jac13}.  
It is also interesting to mention in this connection that the description of such topologically charged nonlinear waves (see e.g., the review \cite{fetter}) and their surprisingly ordered and robust lattices, as well as their role in phenomena as rich and profound as superconductivity and superfluidity were the theme of the most recent Nobel prize in Physics in 2003.

While it is clear that there is an enormous volume of literature on BECs,
as well as on the nonlinear matter-waves present in this setting, in this topical review, we will have to narrow our scope and limit ourselves to the 
study of ``pattern forming dynamical instabilities'' in BECs, i.e., those that give rise to the generation of coherent structures, such as matter-wave solitons and vortex arrays. As we aim to focus on the origin of these patterns--which have been observed/realized and studied in many different physical contexts as well-- 
we will give a (undoubtedly, somewhat subjective) view of the excitement that this new world of ``matter waves'' has brought to diverse disciplines, including atomic physics, nonlinear science, wave physics and nonlinear optics, while providing a perspective of the multitude of intriguing connections and parallels between these. 

To be more specific, we will first examine the {\it modulation instability} (MI) of plane waves, the activation of which, is perhaps the most standard mechanism through which bright solitons and solitary wave structures appear in the context of the ``traditional'' NLS equation (without the external potential). In this case, the plane wave solution of the NLS equation becomes unstable towards the formation of a pattern consisting of a chain (or ``train'') of bright solitons. The demonstrations of MI span a diverse set of disciplines, ranging 
from fluid dynamics \cite{benjamin} (where it is usually referred to as the 
Benjamin-Feir instability) and nonlinear optics \cite{ostrovsky} (see also \cite{hasegawa} for a review) to plasma physics \cite{taniuti68}. In the BEC context under consideration, we aim to present how quasi-$1$D BECs, trapped in regular magnetic traps or optical lattice potentials (in both continuous and discrete settings, as well as in time-dependent cases which are relevant to the BEC dynamics in the presence of external time-dependent fields), can be destabilized giving rise to such coherent matter-wave structures. We will also derive a relevant instability threshold (which will be given in terms of physical BEC parameters, thus being quite relevant for experiments), providing in this way a simple criterion for avoiding the MI. 

On the other hand, as far as higher-dimensional settings are concerned, the most fundamental nonlinear excitations of BECs are vortices, which can be excited by phase imprinting, by stirring or rotating the condensate (as mentioned above), or by the transverse decay of dark solitons (see, e.g., \cite{dark}). The latter mechanism is directly connected to the onset of the {\it transverse modulational instability} (also called simply ``transverse instability'' (TI) or ``snaking instability'' (SI)) of {\it rectilinear} dark solitons, which has been studied extensively in  nonlinear optics (see, e.g., \cite{luther,depysk} for a review), as well as in the BEC context \cite{tibec}. Therefore, the TI is of particular interest here, as it is another ``pattern forming instability''--relevant in higher-dimension BECs--which gives rise to vortex patterns. Here, it is interesting to note that apart from the vortex lattices mentioned above, very robust vortex arrays, in the form of {\it vortex necklaces} composed of vortex-antivortex pairs, can also be formed through the onset of TI of {\it ring} dark solitons (RDS) in BECs \cite{prlrds} (note that relevant patterns, but composed by globally linked vortices of one sign rather than of the vortex-antivortex type, were also predicted to occur in $2$D BECs \cite{crasovan}). In this work, we will discuss how the TI destabilizes the RDS, leading to the formation of vortex-antivortex necklaces. Also, similarly to the case of the MI in $1$D settings, we will derive the TI threshold (in terms of physical BEC parameters), providing in this way a criterion for avoiding the instability. 

The structure of our presentation will be as follows:
\begin{itemlist}
\item In section 3, we will discuss reductions of the $3$D 
GP equation
to lower dimensional settings including 
$1$D and $2$D continuum and lattice ones.
\item In section 4, we will examine the modulational instability in 
continuum 
$1$D settings and demonstrate how it gives rise to solitary waves.
\item In section 5 we will focus on the modulation instability in discrete settings, and in particular in BECs trapped in optical lattices.
\item In section 6, we will turn our attention to $2$D scenaria and 
show how the snaking/transverse instability gives rise to vortex structures.
\item In section 7, after briefly touching upon some of the solitary wave
structures and features analyzed in the BEC setting, we summarize our findings and present some directions for future study.
\end{itemlist}

\section{Reductions to Lower Dimensions}

The genuinely three-dimensional ($3$D) BECs can be considered as approximately $1$D if the nonlinear inter-atomic interactions are weak relative to the trapping potential force in the transverse  directions; then, the transverse size of the condensates is much smaller than their length, i.e., the BEC is ``cigar-shaped'' and can be effectively described by $1$D models 
\cite{GPE1d,VVK}. Similarly, if the transverse confinement is strong along one direction and weak along the others, then for this ``pancake-shaped'' BEC, $2$D model equations are relevant \cite{GPE2d}. 

Close to zero temperature, it is well-known that the $3$D 
GP equation 
\cite{review} accurately captures the dynamics of the condensate. For the cigar-shaped BEC, the model equation is effectively $1$D and can be expressed in the following dimensionless form: 
\begin{eqnarray} 
i \frac{\partial \psi}{\partial t}= -  
\frac{\partial^2 \psi}{\partial x^2} + g |\psi|^2 \psi + V(x) \psi. 
\label{eqn1} 
\end{eqnarray} 
%

In this equation, the normalized macroscopic wave function $\psi$ is 
connected to the original order parameter $\Psi$ through the equation 
$\psi=(\epsilon/\sqrt{2\pi |a|}a_{\perp})^{-1} \exp \left[i\omega_{\perp}t+
i(y^{2}+z^{2})/2a_{\perp}\right]\Psi$, while $t$ and $x$ are respectively measured in units of 
$3/2\epsilon^{2}\omega_{\perp}$
and 
$\sqrt{3}a_{\perp}/2\epsilon$. Here, 
$\omega_{\perp}$ is the confining frequency in the transverse  
direction, $a_{\perp}=\sqrt{\hbar/m \omega_{\perp}}$ is the transverse harmonic-oscillator length and $\epsilon \equiv N |a|/a_{\perp}$ is a small dimensionless parameter, $N$ being the total number of atoms and $a$ the scattering length. Finally, 
$g=\pm 1$ is the renormalized nonlinear coefficient, 
which is positive (negative) for repulsive (attractive) condensates, while  
$V(x)=(1/2)\Omega^{2} x^{2}$ is the magnetic trappping potential, where  
$\Omega=(3/2\sqrt{2}\epsilon^{2}) \cdot (\omega_x/\omega_{\perp})$
with $\omega_x$ being the axial confining frequency. Such a reduction of the $3$D GP equation to $1$D can be done self-consistently using a multiscale expansion based perturbation technique \cite{VVK,VVK2} [with the wave function $\psi$ 
being of order $O(1)$]; for a rigorous derivation, the 
interested reader can refer to \cite{lieb2} and references therein. 

In the same way, the $2$D model for the pancake-shaped condensate assumes the form: 
\begin{eqnarray} 
i \frac{\partial \psi}{\partial t}= -  
\Delta \psi + g |\psi|^2 u + V(r) \psi 
\label{eqn2} 
\end{eqnarray} 
where the roles of the axial frequency $\omega_{x}$ and the spatial variable $x$   
are now played by the radial frequency $\omega_{\perp}$ and $r\equiv \sqrt{x^{2}+y^{2}}$ respectively; also, the normalized variables are connected to the dimensional ones similarly to the $1$D case, but with $\omega_{x}$ ($\omega_{\perp}$) instead of $\omega_{\perp}$ 
($\omega_{z}$). We note that the use of self-consistent reductions of the 
$3$D model to lower-dimensional ones, allows (for different normalizations 
of the wave function) for different choices of the values of coefficients of Eqs. (\ref{eqn1}) or (\ref{eqn2}) (e.g., the coefficient of the kinetic energy term may be set equal to $1/2$, and/or the nonlinearity coefficient $g$ may take values different than $\pm 1$, and so on; see also a relevant discussion in section 6). 

Similar reductions can be performed in the presence of the 
OL potential: In particular, considering an optical lattice 
along the longitudinal direction of a cigar-shaped condensate, 
the GP model of Eq. (\ref{eqn1}) 
is relevant with a potential \cite{ol1,ol2} 
$V(x)=V_0 \sin^2(k x)$,
where $V_{0}$ is the normalized strength of the OL (measured in units of the recoil energy 
$E_{r}\equiv h^2/2m\lambda^{2}$) 
and $\lambda=2\pi/k$ is the inter-well distance in the interference pattern. In the 
$2$D case, the egg-carton potential of the form 
$V(x,y)=V_0 \left[\sin^2(k_x x)+ \sin^2(k_y y)\right]$ 
is the relevant OL potential for the pancake-shaped condensates.

Another useful reduction \cite{wann1,wann2} (see also \cite{tromb,konot}) for periodic (e.g., optical lattice) potentials is the one of the GP equation to the discrete NLS equation (DNLS) \cite{IJMPB,johans}. Starting from 
%
%
Eq. (\ref{eqn1})
and assuming that the potential $V(x)$ is 
periodic, i.e., $V(x+L)=V(x)$, we consider the associated eigenvalue problem
%
\begin{equation}
\frac{d^{2}\varphi _{k,\alpha }}{dx^{2}}+V(x)\varphi _{k,\alpha }=E_{\alpha
}(k)\varphi _{k,\alpha },  \label{eigen}
\end{equation}
where $\varphi _{k,\alpha }$ has  Bloch (Floquet) functions (BF's) 
$\varphi _{k,\alpha}=\exp(ikx)\psi_{k,\alpha}(x)$, with $\psi_{k,\alpha}(x)$ 
periodic with period $L$ ($\alpha$ is an index labeling the energy bands $E_{\alpha}(k)$).
For our purposes it is more convenient to 
use the Wannier functions (WF's) \cite{kohn} instead of the BF's:
\beq
w_{\alpha }(x-nL)=
\sqrt{\frac{L}{2\pi }}\int_{-\pi /L}^{\pi/L}\varphi _{k,\alpha }(x)e^{-inkL}dk. 
\label{DefW}
\label{wfdef}
\eeq
Due to completeness of WF's, any solution of Eq. (\ref{eqn1}) can be expressed in the form
\begin{equation}
\psi (x,t)=\sum_{n\alpha }c_{n,\alpha }(t)w_{n,\alpha }(x),  \label{expan1}
\end{equation}
which after substitution in (\ref{eqn1}) gives
\begin{eqnarray}
i{\frac{dc_{n, \alpha}}{dt}} = 
\sum_{n_1}c_{n_1,
\alpha} \hat{\omega}_{n-n_1,\alpha } +
 g \sum_{\alpha_1, \alpha_2, \alpha_3}\sum_{n_1,n_2,n_3}
  c_{n_1,\alpha_1}^{*} c_{n_2,\alpha_2} c_{n_3,\alpha_3} W^{n n_1 n_2
n_3}_{\alpha \alpha_1 \alpha_2 \alpha_3 },
\label{exact}
\end{eqnarray}
where
\begin{equation}
W^{n n_1 n_2 n_3}_{\alpha \alpha_1 \alpha_2
\alpha_3} = \int_{-\infty}^{\infty}
w_{n,\alpha}w_{n_1,\alpha_1}w_{n_2,\alpha_2}w_{n_3,\alpha_3} dx
\label{overlapp}
\end{equation}
are overlapping matrix elements. Upon suitable decay of the Fourier coefficients and the WF prefactors (which can be systematically checked for given potential parameters), the model can then be reduced to 
\begin{eqnarray}
i {\frac{dc_{n, \alpha}}{dt}}&=&\hat{\omega}_{0,\alpha}c_{n,\alpha}
+ \hat{\omega}_{1,\alpha }\left(c_{n-1,\alpha} +c_{n+1,\alpha}\right)
\nonumber \\
&+& g \sum_{\alpha_1, \alpha_2, \alpha_3}
 W^{nnnn}_{\alpha \alpha_1 \alpha_2 \alpha_3 }  c_{n,\alpha_1}^{*}
 c_{n,\alpha_2} c_{n,\alpha_3}.
\label{i-ii}
\end{eqnarray}
The latter equation degenerates into the tight-binding model \cite{tromb,konot}
\begin{eqnarray}
i \frac{dc_{n,\alpha}}{dt} =\hat{\omega}_{0,\alpha}c_{n,\alpha}
+ \hat{\omega}_{1,\alpha}\left(c_{n-1,\alpha}
+c_{n+1,\alpha}\right)
+g W^{nnnn}_{\alpha \alpha \alpha \alpha}  |c_{n,\alpha}|^2 c_{n,\alpha},
\label{TB}
\end{eqnarray}
if one restricts consideration to the band $\alpha$ only.

We have now laid out the setting: It can consist of the $1$D GP equation with a magnetic trap or an optical lattice potential (or both); it can also be of the form of a $2$D GP equation with a magnetic trap and/or an optical lattice. Finally, it can also consist of genuinely discrete lattice dynamical systems. In these contexts, we now proceed to study some of the pattern forming instabilities.

\section{Modulational Instability in quasi-$1$D BECs}

\subsection{The Case of 
the Untrapped Condensate}

In the case of the untrapped condensate, i.e., for $V(x)=0$, the standard modulational instability (MI) analysis concerns the investigation 
of the stability of the plane wave solutions of the model  
equation (\ref{eqn1}) \cite{benjamin,ostrovsky,hasegawa}. In particular, in the absence of a potential, Eq. (\ref{eqn1}) (which is actually the ``regular'' NLS equation) 
has solutions of the form $\psi(x,t)=\psi_0 \exp[i (k x-\omega t)]$, with $\psi_{0}$, $k$ and $\omega$ being the amplitude, wavenumber and frequency of the plane wave respectively, which are connected to each other  
through the dispersion relation $\omega=k^2 + g \psi_0^2$.

Then, to study the stability of the above mentioned solution, we 
introduce the following ansatz into Eq. (\ref{eqn1})
\beq
\psi(x,t)=(\psi_0 +\epsilon b) \exp[i ((k x-\omega t) +\epsilon w(x,t))],
\eeq
where the amplitude and phase perturbations are assumed to have the form 
$b(x,t)=b_{0} \exp(i (Q x-\Omega t) )$ and $w(x,t)=w_{0} \exp(i (Q x-\Omega t))$
respectively ($Q$ and $\Omega$ are the perturbation wavenumber and frequency, while $b_{0}$ and $w_{0}$ are constants). Then, assuming that the parameter $\epsilon$ is small, we can derive [to order $O(\epsilon)$] the following dispersion relation,
%
\beq
(\Omega-2k Q)^{2}=Q^{2}(Q^{2}+ 2g \psi_0^{2}),
\label{meq}
\eeq
which signifies that, for $g=-1$, wavenumbers $Q<Q_{cr} \equiv \psi_{0} \sqrt{-2g}$ will be modulationally unstable, hence the system is unstable to long-wavelength perturbations of this form. 
Importantly, the modulational instability in the absence of potentials
can only occur in the case of attractive inter-particle interactions ($g=-1$), while it is not feasible for repulsive interactions ($g=+1$).

\begin{figure}[th]
\centerline{\psfig{file=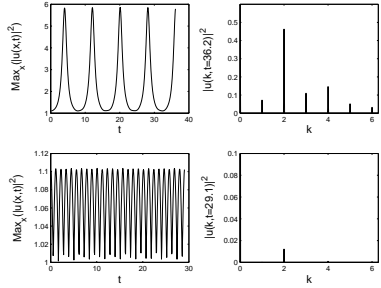,width=5cm}}
\vspace*{8pt}
\caption{The top panel shows the maximum of the solution as a function
of time (top left) and the Fourier space picture at the end of the
numerical experiment (top right) in a modulationally unstable case ($Q=1$).
In contrast, the bottom panels show the same features for a modulationally
stable case ($Q=2$). In both cases, the perturbation $0.05\sin(Q x)$ was added to the uniform solution $\psi_{0}=1$.}
\label{rfig1}
\end{figure}

The manifestations of the MI in numerical experiments are shown in Fig. \ref{rfig1}. 
It can be clearly observed that MI stable wavenumbers, when perturbed,
 only sustain small oscillations (no sidebands are generated in Fourier space), while unstable wavenumbers lead to the generation of {\it soliton trains} i.e., arrays of large amplitude excitations and the creation of sidebands in momentum space.

Notice that one can take an alternative route to establishing MI, recently introduced in \cite{zoi}. Particularly, let us consider 
the ansatz,
\beq
\psi=[\psi_0+a(t) \exp(i \phi_a(t)) \exp(i Q x)+b(t) \exp(i \phi_b(t)) 
\exp(-i Q x)] \exp\left(i(k x-\omega t) \right)
\label{seq8}
\eeq
presenting, as previously, a modulation of the plane wave, with $a(t)$, $b(t)$ being time-dependent functions. Then, Eq. (\ref{seq8}) is introduced not in Eq. (\ref{eqn1}), but instead in the field-theoretic Lagrangian
(from which the equation can be derived), namely,
\begin{eqnarray}
L=\int_{-\infty}^{\infty} \left[ \frac{i}{2} \left(\psi^{\ast} \psi_t-
\psi \psi_t^{\star} \right) -
|\psi_x|^2- \frac{1}{2} g|\psi|^4 \right]~ dx.
\label{seq7}
\end{eqnarray}
Deriving the Euler-Lagrange equations and keeping the terms to leading order [i.e., to $O(a)$], we obtain $a=b$, $\phi=\phi_a+\phi_b$, and
\begin{eqnarray}
\dot{a}&=&C_2 a \sin(\phi)
\label{seq18}
\\
\dot{\phi}&=&(C_1+C_3)+2 C_2 \cos(\phi),
\label{seq19}
\end{eqnarray}
where $C_1 = -g \psi_0^2-Q^2- 2Qk$, $C_2 = -g \psi_0^2$,
and $C_3 = -g \psi_0^2-Q^2+ 2Qk$ are constant prefactors.
Then, one can solve the ensuing ordinary differential
equations for $a$ to obtain:
\begin{itemlist}
\item When $Q^2+2g \psi_{0}^{2}>0$, the solution of Eq. (\ref{seq18}) is
\begin{eqnarray} 
a(t) \sim \sqrt {1+\frac{2 g \psi_0^2}{Q^2} \sin 
\left(\sqrt{(Q^2+2 g \psi_0^2)Q^2}t \right)^2}.
\label{seq21}
\end{eqnarray}
\item When $Q^2+2 g \psi_0^2<0$, the solution of Eq.(\ref{seq18}) is
\begin{eqnarray} 
a(t) \sim \sqrt{1-\frac{2 g \psi_0^2}{Q^2} \sinh
\left(\sqrt{(-2 g \psi_0^2-Q^2)Q^2} t \right)^2},
\label{seq22}
\end{eqnarray}
\end{itemlist}
and hence instability will arise in the latter case (for $g=-1$ and $Q<Q_{cr} \equiv \psi_{0} \sqrt{2}$), while it will be absent in the 
former one (for $Q>Q_{cr}$). 
 It is interesting to note that the degenerate critical case of $Q=Q_{cr}$
is also unstable, however the growth in the latter is polynomial rather than
exponential; in particular, in the latter case $a(t) \sim \sqrt{1+4 g^2 
\psi_0^2 t^2}$.
We should remark here that this technique can be applied to any 
system 
exhibiting a Lagrangian structure.

\subsection{
The Effect of Linear or Quadratic Potentials}

While the standard analysis of the MI is instructive, in the more realistic case of trapped 
BECs, the interesting variation from the regular NLS equation lies in the existence of external potentials in the present context.

The simplest such example (also one that can be fully analyzed) is the presence of a linear (e.g., gravitational \cite{kasevich}) potential $V(x)={\cal E} x$ in Eq. (\ref{eqn1}) 
\cite{mizoi,migeorge}. In this setting, the NLS is well-known to maintain its integrable character \cite{liu}. Hence, in some sense, we expect that the modulational instability results/conditions will not be modified in this case.

The simplest way to illustrate that is by means of the so-called ``Tappert transformation'' \cite{liu} 
\begin{eqnarray}
\psi(x,t)=v(\eta,t) \exp \left(-i {\cal E} x t-\frac{1}{3} i {\cal E}^2 t^3\right),
\label{req9}
\end{eqnarray} 
where $\eta= x+ {\cal E} t^2$. This transformation brings Eq. (\ref{eqn1}) back into the form of the regular NLS equation, without the external potential, in which the MI condition given in the previous subsection applies.

The physically more interesting case is the one concerning a quadratic potential, say $V(x)={\cal K} x^2$, with $\cal K$ being in general a function of time. This case, however, is 
less mathematically tractable. One trick 
that can be used in this case is the so-called 
lens-type transformation \cite{sulem} 
of the form:
\begin{equation}
\psi(x,t)=\ell^{-1} \exp \left[i f(t) x^2\right]v(\zeta,\tau),
\label{req11}
\end{equation}
where $f(t)$ is a real function of time, $\zeta= x/\ell(t)$ and 
$\tau=\tau(t)$. 
To preserve the scaling we choose \cite{sulem}
\begin{eqnarray}
\tau_{t}=1/\ell^2.
\label{req13b}
\end{eqnarray}
To satisfy the resulting equations, we then demand that:
\begin{eqnarray}
-f_{t}=4f^2+{\cal K}(t)
\label{req14a}
\\
-\ell_{\tau}/\ell+4 f \ell^2=0.
\label{req14b}
\end{eqnarray}
Taking into account Eq. (\ref{req13b}), the last equation can be solved:
\begin{equation}
\label{req14c}
\ell(t)=\ell(0)\exp\left(4\int_0^tf(s)ds\right).
\end{equation}

Upon the above conditions, the equation for $v(\zeta,\tau)$ becomes
(for $g=-1$)
\begin{equation}
iv_{\tau}+v_{\zeta\zeta}+|v|^2 v-2 i\gamma v =0,  
\label{req15}
\end{equation}
where 
$f \ell^2=\gamma,$
and generically $\gamma$ is real and depends on time. Thus we retrieve a NLS equation with an additional term, which represents either growth (if $\gamma>0$) or dissipation (if $\gamma<0$). 
 
Eq. (\ref{req15}) leads to an explicit, spatially homogeneous solution of
the form: 
$$
\psi = \frac{A_0}{\ell(t)}  
    \exp \left[i f(t) x^2 + i 
\frac{kx}{\ell(t)} - i k^2 \tau(t) \right] 
\times
$$
\begin{eqnarray}
\exp \left[ \Gamma(\tau)+ i A_0^2 \int_0^{\tau} \exp\left(2 \Gamma(s) \right) ds + i \theta_0 \right],
\label{exact1}
\end{eqnarray}
where $\Gamma(\tau)= 4 \int_0^{\tau} \gamma(s) ds$, and $A_0$ and $\theta_0$ are arbitrary real constants.

\begin{figure}[th]
\centerline{\psfig{file=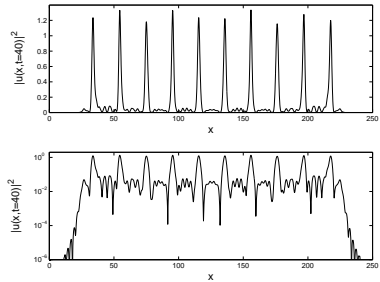,width=5cm}}
\vspace*{8pt}
\caption{The result of a numerical simulation in a regular (top panel)
and semi-log (bottom panel) plot for the wavefunction in a potential
decaying $\sim (t+t^{*})^{-2}$. A soliton train is clearly formed.}
\label{rfig2}
\end{figure}

A particularly interesting case is that of $\gamma$ constant. 
Then, from the system of equations
(\ref{req13b})-(\ref{req14b})
and (\ref{req15}), 
it follows that $\cal K$ must have a specific form and, thus, the functions $f$, $\ell$ and 
$\tau$ can be determined accordingly. In fact, the system (\ref{req13b})-(\ref{req14b})
and (\ref{req15}) with $\gamma$ constant has as its 
solution
${\cal K}(t)=(1/16)(t+t^{*})^{-2}$ ($t^{*}$ is an arbitrary constant),
$f(t)=(1/8)(t+t^{*})^{-1}$, 
$\ell(t)=2 \sqrt{2 \gamma} \sqrt{t+t^{*}}$ and
$\tau(t)=(1/8 \gamma)\ln (\frac{t+t^{*}}{t^{*}})$.
This case can be analyzed completely and for details the interested reader is referred to 
\cite{hasegawa,lisak}. 
Importantly, relevant numerical simulations corresponding to 
this case, reveal the formation of solitonic patterns, and in particular 
trains of matter-wave bright solitons, 
such as the one of Fig. \ref{rfig2} in the case of this 
time decaying potential.

\begin{figure}[th]
\centerline{\psfig{file=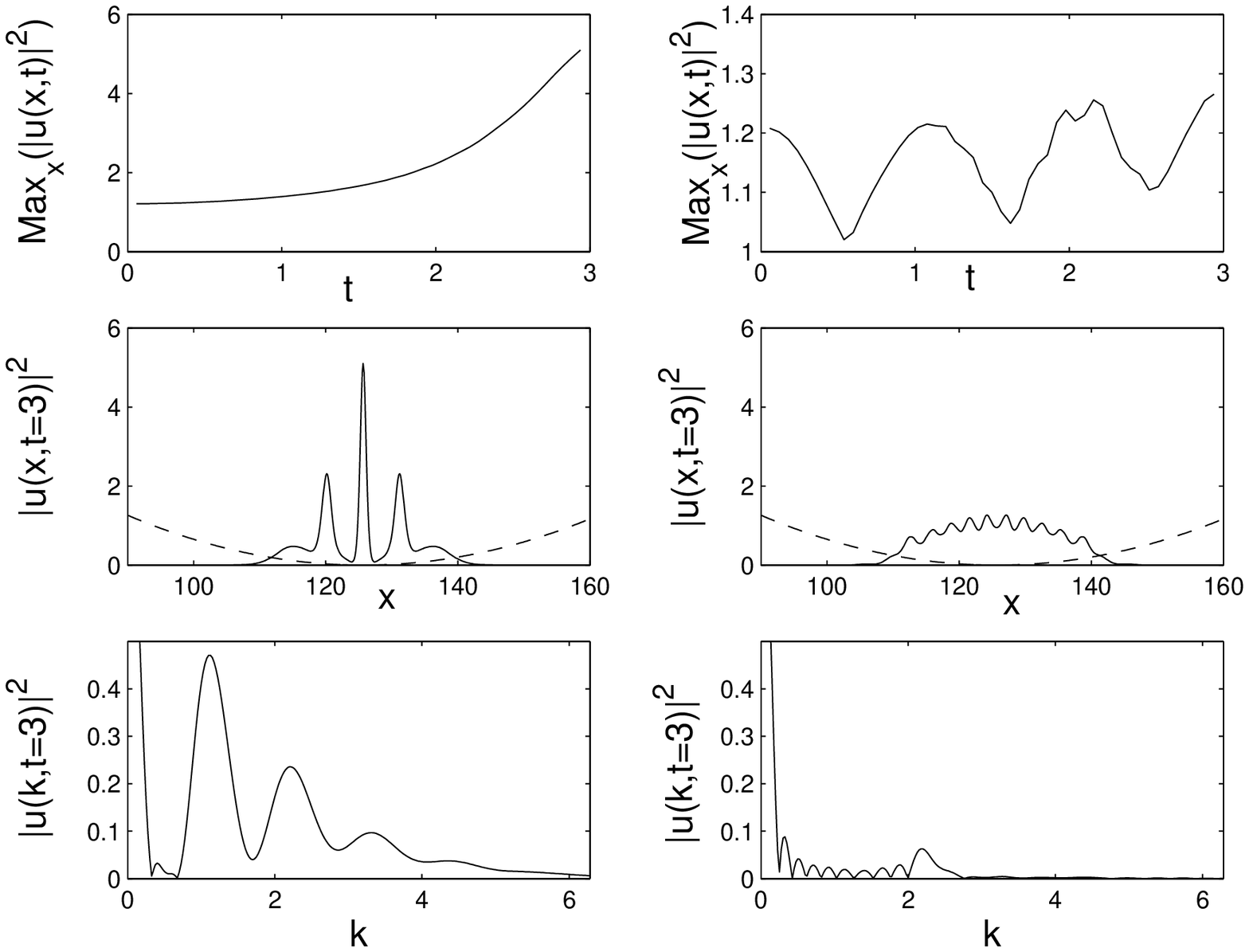,width=5cm}}
\vspace*{8pt}
\caption{The top panel shows the time evolution of the wavefunction
maximum in a (previously) modulationally unstable (left) and stable 
(right) case. The middle panel shows the respective spatial profiles
at $t=3$ and the bottom the corresponding Fourier pictures. Noticeably,
the instability starts to develop even in the previously modulationally
stable case.}
\label{rfig3}
\end{figure}

In the general case, $\lambda$ must be time dependent (i.e., $\lambda \equiv \lambda(t)$).
Here one cannot directly perform the MI analysis and is essentially limited to the realm of numerical simulations. Figure \ref{rfig3} shows such an example in the presence of the magnetic trap. It can be seen that modulationally unstable wavenumbers result in the formation of soliton wavetrains modulated by the harmonic potential. This is a picture qualitatively 
reminiscent of the experimental outcome of the modulational instability in attractive condensates illustrated in \cite{bright1}. On the other hand, if we initialize the system with a modulationally stable configuration, then 
we note that, interestingly, the instability
manifests itself in the latter case as well.
The reason for the occurrence of MI in both cases is that the dynamics of the 
potential in Eq. (\ref{eqn1}) mixes the wavenumbers of the original perturbation and eventually results in the excitation of modulationally 
unstable ones. However, this only happens 
later (because firstly the modulationally 
unstable $Q$'s need to be excited) and with a smaller amplitude in this case. 

We should note here that the experimental results of \cite{bright1} have significantly increased the interest in the modulational instability of NLS equations with potentials
and various additional terms emulating the experimental conditions (see also below). Such examples can be found e.g., in the works of \cite{spl,khawaja,leung,carr}.

\subsection{
Time-Dependent Settings: Feshbach Resonance Management}

Feshbach resonances \cite{feschb,feshb1,feshb2} have recently become 
a popular tool for manipulating 
the dynamical evolution of BECs. They have been used to induce or prevent collapse type phenomena \cite{feshb1,feshb3}, to study shock waves \cite{vvkvpg}, and to create condensates with tunable interactions (e.g., attractive BECs starting from repulsive ones) more generally \cite{marte}. The resonance allows one to adjust and essentially tune the scattering length 
[$g$ in the context of Eq. (\ref{eqn1})] at will 
--and even change its sign-- by using a 
resonantly tuned ac magnetic field.

To generate breathing 
solitons in one spatial dimension (in the presence of the magnetic trap  
\cite{VVK,FRM,dep}, or in the quasi-discrete setting created by an optical-lattice potential  
\cite{abdul2}), as well to avoid their collapse in two dimensions \cite{feshb3}, an experimentally realizable protocol has been recently proposed, in the form of the so-called ``Feshbach Resonance Management'' (FRM). 

The FRM scheme can be modelled (in the mean-field approximation) in the framework of the GP equation with the coefficient in front of the nonlinear term being a periodic function of 
time. In Ref. \cite{FRM}, the periodic function was taken to be a piece-wise constant one, periodically jumping between positive and negative values. The same model may also be realized in terms of nonlinear optics, where it applies to a medium composed of alternating layers with opposite signs of the Kerr nonlinearity \cite{Isaac}. FRM resembles the 
\textit{dispersion-management} (DM) scheme, well-known in fiber optics (see, e.g., 
\cite{Progress} and \cite{turits} for relevant reviews), which assumes that 
long fiber links are formed by a concatenation of nonlinear 
fibers with opposite signs of the group-velocity 
dispersion periodically alternating, 
thus realizing a system that supports robust breathing 
solitons as information carriers.

This context motivates the study of the MI in a setting where Eq. (\ref{eqn1}) has time-dependent coefficients. In particular, the most general such setting would be:
\begin{equation}
i\psi_{t}=-D(t)\psi_{xx}+g(t)|\psi|^{2}\psi+ V(x)\psi.  
\label{fmeq1}
\end{equation}
Assuming for simplicity the absence (or weak spatial dependence) of the potential, 
the plane-wave solution to Eq. (\ref{fmeq1}) is 
\begin{equation}
\psi_{o}=A_{0}\exp \left[ i \left(kx-k^{2}\int_{0}^{t}D(s)ds-A_{0}^{2}\int_{0}^{t} g(s)ds\right)\right].
\label{pw}
\end{equation}
Note that the arbitrary amplitude $A_0$ in Eq. (\ref{pw}) can be rescaled to $A_{0}\equiv 1$. 
Then, we 
seek, as in the previously studied cases, solutions of Eq. (\ref{fmeq1}) incorporating a modulation that perturbs the above mentioned plane wave, having now the form
\begin{equation*}
\psi=\psi_{o}\left[ 1+\epsilon w(t)\cos (Q x)\right] ,
\end{equation*}
where $\epsilon$ and $Q$ are the amplitude and wavenumber of the perturbation. This way, we derive the following (linearized) equation for the real part, $w_r \equiv Re(w)$, of the perturbation,
\begin{equation}
\ddot{w}_{r}=\dot{D}D^{-1}\dot{w}_{r}-Q^{2}D(t)\left[ Q^{2}D(t)+2 g(t)\right]w_{r},  
\label{fmeq2}
\end{equation}
where the overdot stands for $\mathrm{d/dt}$. There are several special cases of this equation that were previously studied. In the case of DM (i.e., for $D=D(t)$ and $a(t)\equiv \mathrm{const.}$), the MI analysis was performed in \cite{doran}. On the other hand, in the
FRM context, for $D\equiv 1/2$ and time-periodic $g(t)$, Eq. (\ref{fmeq2}) is a Hill equation that was considered in \cite{FRM}, while the more specialized case of $g(t)=1+2\alpha \cos (\omega t)$ leads to the Mathieu equation that was studied, in this context, in \cite{stal}.

Motivated by the FRM scheme proposed in \cite{FRM},
Ref. \cite{scripta} 
discusses the special case of a piecewise-constant time-dependent $g$ alternating between $g_{1}$ (for $0<t\le \tau$) and 
$-g_{2}$ (for $\tau<t\le T$) as mentioned above. 
In this case, first, we define 
$s_{1}^{2}=Q^{2}(Q^{2}/4+g_{1})$ and $s_{2}^{2}=Q^{2}(g_{2}-Q^{2}/4)$, which assumes 
$Q^{2}<4g_{2}$. Then, seeking for a solution in accordance with Bloch's theorem \cite{Kittel} within one period and using the appropriate continuity conditions for the integration constants, one can obtain a stability condition. This procedure (similar to the one used for identifying the allowed/forbidden bands for the Kronig-Penney model in quantum mechanics
\cite{Kittel}) leads to the determinant condition
for the eigenfrequency (Floquet multiplier) 
$\Omega $ of the solution: 
\begin{equation}
\cos (\Omega T)=-\frac{s_{1}^{2}-s_{2}^{2}}{2s_{1}s_{2}}\sin (s_{1}\tau
)\sinh [s_{2}(T-\tau )]+\cos (s_{1}\tau )\cosh [s_{2}(T-\tau )]\equiv F(Q).
\label{hill7}
\end{equation}
If the above condition $Q^{2}<4g_{2}$ does not hold, we redefine $\tilde{s}
_{2}=\sqrt{Q^{2}(Q^{2}/4-g_{2})}$, and obtain, instead of Eq. (\ref{hill7}), 
\begin{equation}
\cos (\Omega T)=-\frac{s_{1}^{2}+\tilde{s}_{2}^{2}}{2s_{1}\tilde{s}_{2}}\sin
(s_{1}\tau )\sin [\tilde{s}_{2}(T-\tau )]+\cos (s_{1}\tau )\cos [\tilde{s}
_{2}(T-\tau )]\equiv \widetilde{F}(Q).  \label{hill8}
\end{equation}

By examining the function $\left\vert F(Q)\right\vert $ or $\left\vert 
\widetilde{F}(Q)\right\vert $, defined in Eqs. (\ref{hill7}) and 
(\ref{hill8}), and comparing it to $1$, we can find whether there is a real
eigenfrequency for a given perturbation wavenumber $Q$, or it belongs to a
\textquotedblleft forbidden zone\textquotedblright, which implies the MI.

\begin{figure}[th]
\centerline{\psfig{file=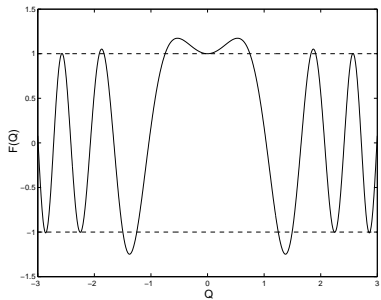,width=5cm}}
\vspace*{8pt}
\caption{The plot of the function $F(Q)$ defined by Eqs. (\ref{hill7}) and 
(\ref{hill8}) for $g_1=g_2=0.3$ and $\protect\tau= T/4= 1$. When the function
satisfies $|F(Q)| \leq 1$, the perturbation of wavenumber $Q$ is
modulationally stable, while for $|F(Q)|>1$, it is modulationally unstable.}
\label{rfig4}
\end{figure}

A typical example of results produced by this analysis is shown in Fig. \ref
{rfig4}, for $g_{1}=g_{2}=0.3$ and $\tau =T/4=1$. 
In general, we have found that the number and widths of the 
instability windows increase as long as the 
mean value $\bar{g}\equiv \lbrack g_{1}\tau
-g_{2}(T-\tau )]/T$ of the scattering length gets large and negative, i.e., when the BEC is
``very attractive'' on the average. 

In \cite{scripta}, the results of the theoretical predictions for the MI were tested against direct numerical simulations of the original PDE and the agreement between the two was found to be generic. Furthermore, the case of the presence of a weak magnetic trap was also examined.

\subsection{Avoiding the Modulational Instability}

In all of the above settings, we examined the conditions for inducing a modulational instability, in order to form solitary wave patterns due to its presence. However, in other contexts, it may be desirable to avoid such dynamical manifestations and have robust structures that are not subject to MI. In such cases, a simple criterion, involving physical parameters of a trapped condensate, for the suppression (or the onset) of the MI would be useful, especially for relevant experiments. As proposed recently in \cite{andrea}, one can derive such a criterion in terms of the magnetic trap strength (which controls the size and the number of atoms of the condensate).

In particular, 
as discussed in section $4.1$, the modulational instability 
sets in for perturbation wavenumbers below a critical one $Q < Q_{cr} \equiv \sqrt{2} \psi_0$ [we set $g=-1$ in the framework of Eq. (\ref{eqn1})]. 
This can be equivalently interpreted as follows: when length scales 
\begin{eqnarray} 
\lambda > \lambda_{cr} \equiv \frac{\sqrt{2}\pi}{\psi_0} 
\label{eqn3} 
\end{eqnarray} 
become ``available'' to the system, then the modulation over these scales leads the plane wave solution to instability.  
 
However, in the presence of the magnetic trap, there is a characteristic scale set by the trap, namely the BEC axial size, $\lambda_{BEC}$, which depends on the trapping frequency $\Omega$. When $\lambda_{BEC} < \lambda_{cr}$, 
suppression of the modulational instability is expected. To estimate $\lambda_{BEC}$ in a specific setup, one can examine e.g., a protocol relevant to the recent experiments reported in \cite{bright1} (to which we referred earlier). 
In particular, 
the initial configuration consists of a $1$D repulsive condensate, whose ground state wave function can be derived in the framework of the so-called 
Thomas-Fermi (TF) approximation 
\cite{review};
subsequently the interaction is made attractive, by taking advantage
of the Feshbach resonance. Therefore, in this situation the length scale of the condensate can be calculated in the TF approximation 
as $\lambda_{BEC} \approx 2\sqrt{2\mu}/\Omega$, where $\mu$ is the chemical potential. Therefore, if 
\begin{eqnarray} 
\Omega> \Omega_{cr} = \frac{12}{\pi^3} \approx 0.387 
\label{eqn5} 
\end{eqnarray} 
(where we took into regard that 
$\mu=(3/2)^{1/3}\Omega^{2/3}$, resulting from the normalization 
condition of the wave function $\psi$), then the 
trapping conditions are ``engineered'' in such a way that 
the modulational instability cannot manifest itself. Such stringent 
trapping conditions were indeed found to preclude the occurrence of 
the MI in the numerical experiments of \cite{andrea}. 

\section{Modulational Instability in BECs trapped in Optical Lattices}

\subsection{Discrete Modulational Instability--Theory and Experiment} 

Another interesting manifestation of the modulational instability has recently been theoretically predicted and
experimentally verified in the context of BECs trapped in an optical lattice.
The latter setting, as discussed in the section $3$, can be
studied (under appropriate conditions) 
in the framework of the tight binding
model of Eq. (\ref{TB}). Hence, we will examine the instability
in the genuinely discrete context to also illustrate how the relevant
conditions are modified by discreteness.

\begin{figure}[th]
\epsfig{figure=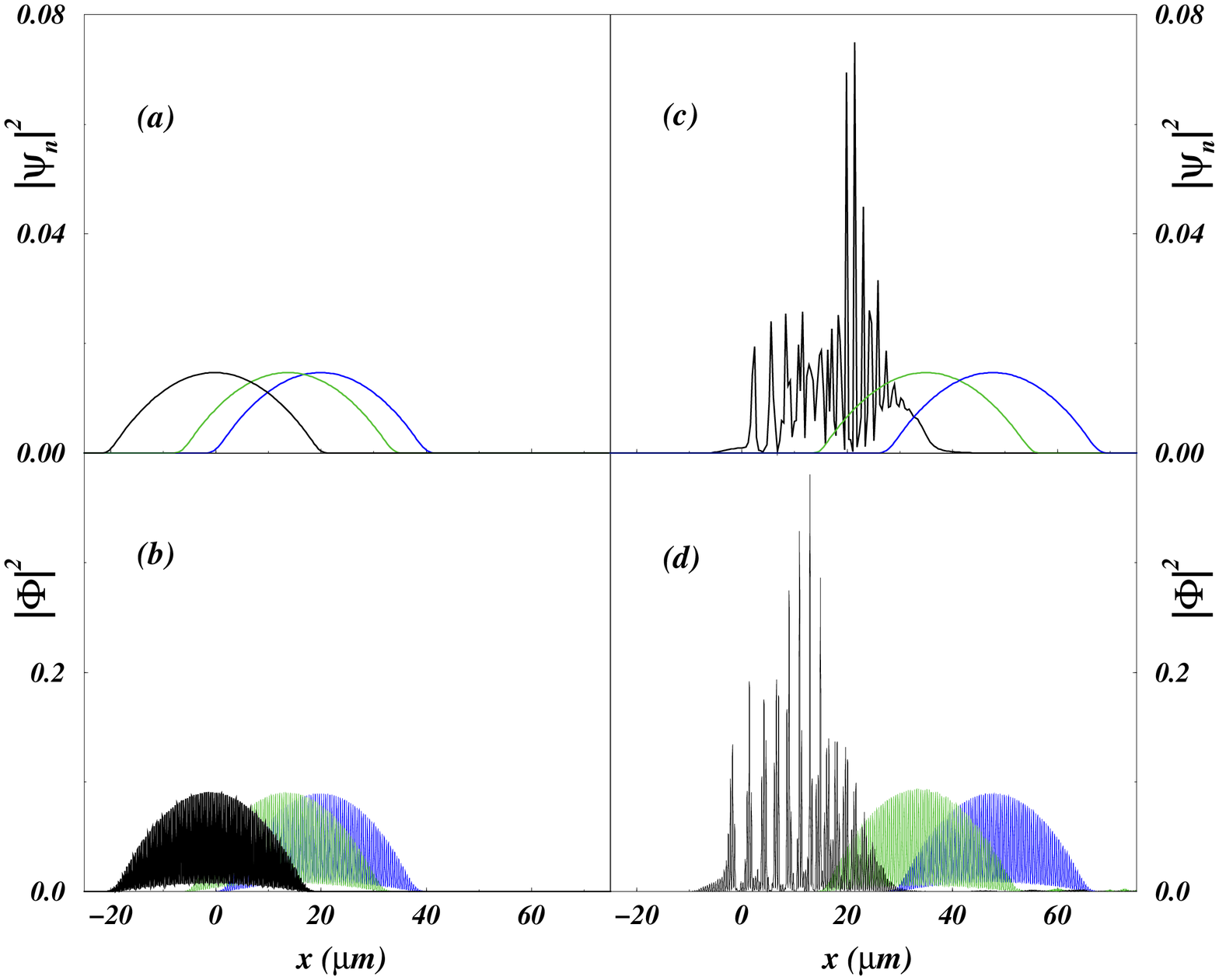,height=2.4in,angle=270}
\epsfig{figure=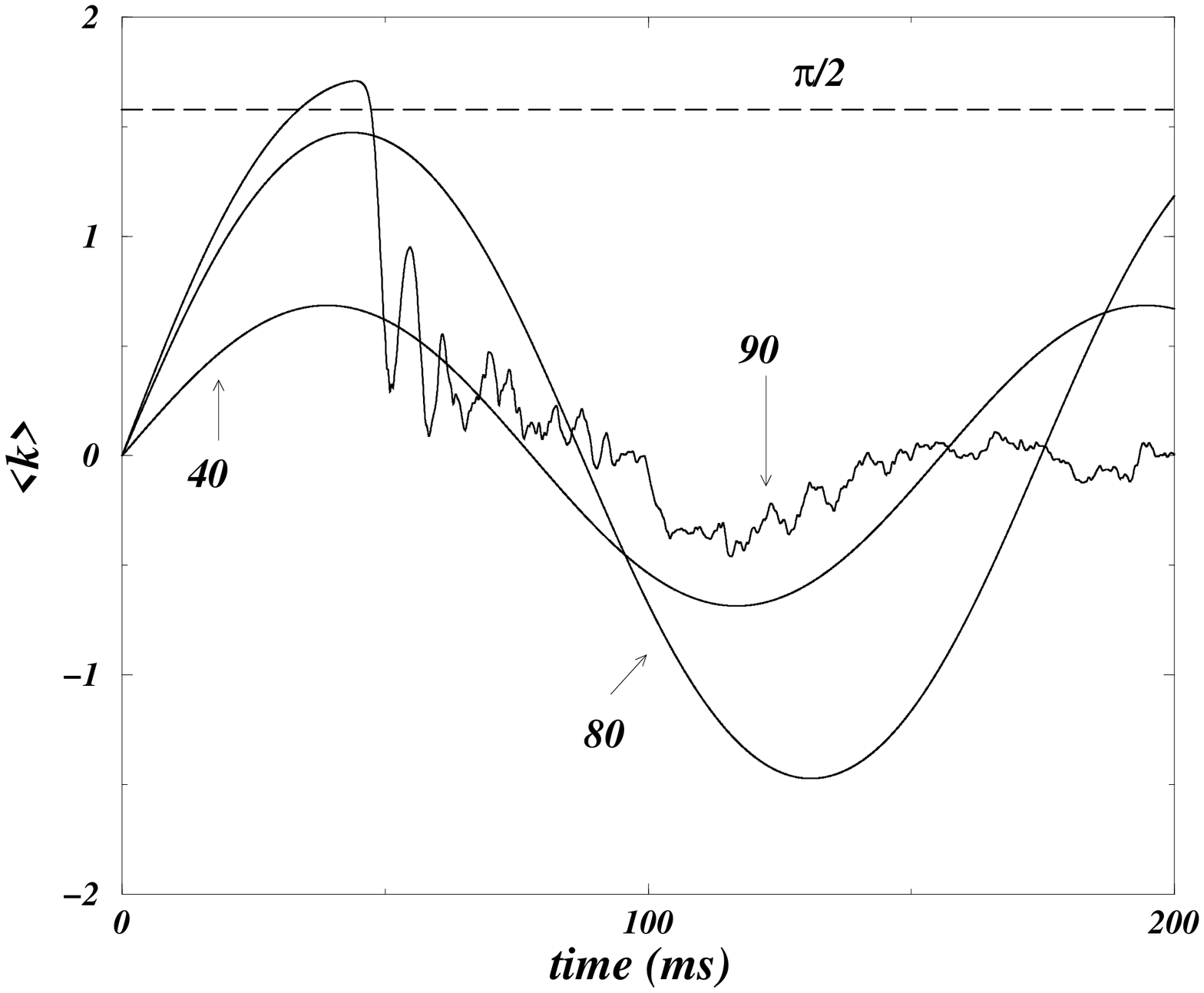,height=2.4in,angle=270}
\vspace*{8pt}
\caption{Left panel: the density $|\psi_n|^2$ at different times 
$0,20,40 ~ ms$ (from 
the right to the left of the figure) for initial 
displacements of (a) $80$ and (b) $100$ sites, respectively 
below and above the critical displacement; 
Right panel: plot of the quasi-momentum $<k>$ vs. time for 
three different initial displacements: $40$, $80$ 
and $90$ sites. The critical displacement is $\xi_{cr} \approx 84$ 
sites. It is clearly seen that when $<k>$ reaches $\pi/2$ 
(i.e., for an initial displacement greater than $\xi_{cr}$), 
the MI is activated.}
\label{rfig5}
\end{figure}

\begin{figure}[th]
\centerline{\psfig{file=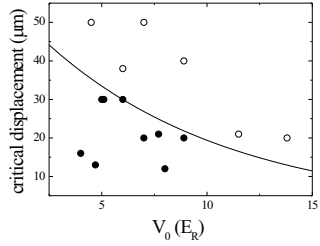,width=5cm}}
\vspace*{8pt}
\caption{Comparison
of experimental results with Eq. (\ref{cr-displ}). Filled circles
represent coherent oscillations, empty circles denote pinned motions;
continuous line results illustrate the critical line of Eq. (\ref{cr-displ}).}
\label{rfig6}
\end{figure}

The relevant discrete reduction of the equation can be re-written
in the general form:
\beq
i  \hbar \frac{\partial \psi_n}{\partial t} = - D
(\psi_{n-1}+\psi_{n+1}) + (V_n+ g \mid \psi_n \mid ^2)\psi_n,
\label{discrete}
\eeq
where $\psi_n$ is the normalized wavefunction of the condensate in the $n$-th site of the array, while the constant $D$ describes the macroscopic tunneling rate between adjacent sites. 
In the context of Eq. (\ref{discrete}), for weak harmonic potentials (such that we can consider $V_n=0$ for the fraction of the lattice considered) the MI analysis can be carried out 
\cite{peyrard} similarly to the continuum case. In particular, taking into regard that the plane wave solution of Eq. (\ref{discrete}) 
has the form $\psi_{n}=\psi_{0}\exp[i(kn-\omega t)]$ 
(with $\omega=-2D\cos(k)+g|\psi_{0}|^{2}$) and performing the stability 
analysis along 
the lines of section $4.1$ with a perturbation 
$\sim \exp(iQn)$, the following condition is readily derived:
\beq
\left(\Omega - 2 D \sin(k) \sin(Q) \right)^2  =
8 D \cos(k) \sin^2({Q \over 2}) \left[2 D \cos(k) \sin^2({Q \over 2}) + 
 g |\psi_0|^2  \right].
\label{condition}
\eeq
The 
crucial difference of this equation from its continuum analog (cf. Eq. (\ref{meq})), identified in \cite{peyrard} and in the context of BECs later in \cite{prl2002}, was the existence of
unstable wavenumbers, even for the case of $g>0$ i.e., for repulsive condensates. In particular, it was appreciated that should wavenumbers $\pi/2< k < 3 \pi/2$ be excited in the
repulsive system, it will then have the potential for instability provided that 
$g |\psi_0|^2 > - 2 D \cos(k) \sin^2(q/2)$ (i.e., for a sufficiently large amplitude excitation). Hence the key to the instability is to excite the modulationally unstable wavenumbers larger than $\pi/2$. Since the experiments are conducted in the presence of a magnetic trap, an indirect way to excite such wavenumbers was suggested in \cite{prl2002} and
subsequently experimentally implemented in \cite{njp2003}. In particular, if the magnetic trap is instantaneously displaced, then the BEC will have to move towards the new center of the 
magnetic trap. This will excite a quasi-momentum (i.e., a wavenumber) which is small if the displacement of the center of the trap is small, but can become larger than $\pi/2$, for sufficiently large initial displacements. By using the Josephson equations 
\beq
\hbar \frac{d}{dt}\xi(t) = 2D ~ \sin{\Delta \phi(t)}   
\eeq
\beq
\hbar \frac{d}{dt}\Delta \phi(t)= - 2 ~ \Omega ~ \xi(t),
\eeq
for the collective coordinates of $\xi=\sum_j j n_j$ (center of mass) and 
$\phi_{j+1}(t)-\phi_j(t) = \Delta \phi(t)$ (quasi-momentum), where $\Omega=(1/2)m\omega_{x}^{2}(\pi/k)^{2}$ is the relevant trap frequency, one can analyze the BEC motion. In particular, it can be obtained \cite{prl2002} that in order for $\Delta \phi$ to become 
$\pi/2$, a displacement of 
\beq
\xi > \xi_{cr}=\sqrt{\frac{2D}{\Omega}}.
\label{cr-displ}
\eeq 
should be applied to the center of the trap. This theoretical prediction was tested in the numerical experiments of \cite{prl2002}, both in the discrete equation and in the continuum one with the periodic potential, yielding very good agreement with the numerical findings (for details see Fig. \ref{rfig5}).
The theoretical results of \cite{prl2002} were also tested directly in experiments of the Florence group, published in \cite{njp2003}. The comparison and good agreement between the experimental results and the theoretical predictions can be summarized in Fig. \ref{rfig6}.

We should also note that a different approach to the modulational instability
in optical lattices was recently proposed in \cite{nicolin}, where it was
observed that the merging of the regular plane wave branch with
an unstable, period-doubled branch of states is responsible for
the onset of the instability. It would be interesting to examine
whether this type of scenario can carry over to different settings.

\subsection{Modulational Instability in a Time-Dependent Optical Lattice}

The context of the optical lattice lends itself to an interesting variant of the 
FRM scheme discussed above. In particular, for example, the heights of the interwell barriers of the optical lattice are proportional to the intensity of the lasers and can be easily modulated periodically in time. This induces an oscillating tunneling amplitude of the condensates between adjacent wells, as well as an oscillating interaction energy of the condensates trapped in each well. This possibility along with the more conventional FRM scheme presented above motivated the recent study of \cite{raptiii} where the coefficients
of Eq. (\ref{discrete}) were allowed to vary in time according to the form:
\begin{equation}
i \dot{\psi}_{n}=-D(t)(\psi_{n+1}+\psi_{n-1})+ g(t) |\psi_{n}|^{2} 
\psi_{n},
\label{deq1a}
\end{equation}
Since one of the two time dependences can be absorbed in a rescaling of time, Eq. (\ref{deq1a}) was examined in \cite{raptiii} with $D(t)\equiv 1$; additionally, in the same work, a periodic variation of the form $g(t)=1+\epsilon \sin(\omega t)$ was assumed for the time-dependent nonlinearity  coefficient $g(t)$.

\begin{figure}[th]
\centerline{\psfig{file=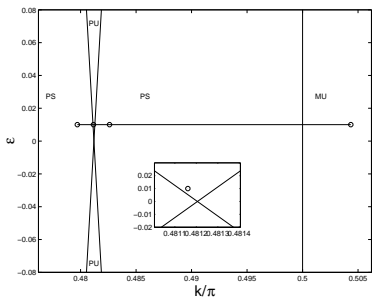,width=5cm}}
\vspace*{8pt}
\caption{The plane of ($\epsilon,k$) where the region of
the parametric instability represented by Eq. (\ref{deq11}) is
shown by two straight lines. Wavenumbers between the two
``stability limits'' will be (to leading order) parametrically unstable.}
\label{rfig7}
\end{figure}

To examine MI in this context, we first note that the plane wave solution of Eq. (\ref{deq1a}) reads:
\begin{equation}
v_{n}(t)=e^{i (k n + 2 \cos(k)t-\int_{0}^{t} g(s)ds)}.
\label{deq5}
\end{equation}
Then, we consider a perturbation ansatz of the following form:
\begin{equation}
\phi_{n}(t)=v_{n}(t) \left[1+\tilde{\epsilon} \left(\alpha(t) e^{i Q n}+\beta(t) e^{-i Q n} \right)\right],
\label{deq6}
\end{equation}
where $Q$ is the perturbation wavenumber and $\alpha(t)$, $\beta(t)$ are 
complex, time dependent fields. The resulting equation for $\alpha$ is then of the form:
\begin{eqnarray}
\hspace{-25mm}
&\ddot{\alpha}+ \left(2 i B-\epsilon \omega \cos(\omega t) \right) \dot{\alpha}& \nonumber \\
&+ \left[A^{2}-B^{2}+2 A+\epsilon (2 A \sin(\omega t)-i \omega (A+B) 
\cos(\omega t) \right]\alpha=0&.
\label{deq8}
\end{eqnarray}
where $A=2 \cos k (1-\cos Q)$ and $B=2 \sin k \sin Q$ are constants depending only on the wavenumber $k$ and the perturbation wavenumber $Q$. To examine this periodic coefficient equation (for possible parametric resonances), a multiple scales technique \cite{bender} was implemented in \cite{raptiii}, using 
$k=k_{0}+ \epsilon k_{1}+O(\epsilon^{2})$.
It was thus found that 
\begin{equation}
k_{1}=\pm \frac{-1+ \sqrt{1+\frac{1}{4}\omega^2}}
{4 \sqrt{1+\frac{1}{4}\omega^2} (1-\cos{Q}) \sin{k_{0}}}.
\label{deq10}
\end{equation}
Therefore, the boundaries for parametric instabilities on the $(k,\epsilon)$ plane can be identified as:
\begin{equation}
\pm \epsilon=\frac{4 \sqrt{1+
\frac{1}{4}\omega^2} (1-\cos{Q}) \sin{k_{0}}}
{-1+\sqrt{1+\frac{1}{4}\omega^2}} (k-k_{0}),
\label{deq11}
\end{equation}
and the result is demonstrated in diagrammatic representation of 
Fig. \ref{rfig7}. The points in the figure represent examples of 
wavenumbers in the different regions of parameter space. In \cite{raptiii} the numerical results indicated good agreement with the theoretical prediction except for the region between
the parametrically and modulationally unstable intervals (where higher-order parametric resonances may set in). Finally, the difference between the manifestations of the MI and of the 
parametric instability which can occur in this setting, is that the former has a considerably larger growth rate and hence develops much earlier (as was analytically calculated and numerically verified).

\section{Transverse (Snaking) Instability in quasi-$2$D BECs}
\subsection{Transverse Instability of Dark Solitons}

We now turn our attention to higher-dimensional settings and discuss 
a similar, long-wavelength, instability in the $2$D setting, relevant to repulsive pancake-shaped BECs. In particular, we will consider the transverse modulational instability of dark solitons, which results to a transverse snake deformation of the soliton (hence the name ``snaking instability'' (SI)) \cite{luther,depysk}, causing the nodal plane to decay into vortex pairs. 

Before proceeding to demonstrate how the SI of dark solitons may lead to robust vortex patterns (vortex necklaces) in BECs, it is relevant to recall some qualitative results concerning the SI. First, it is convenient to renormalize Eq. (\ref{eqn2}) (for $g=+1$) to put it into the form \cite{prlrds}, 
\beq
i \frac{\partial \psi}{\partial t}= -  
\frac{1}{2} 
\Delta \psi + |\psi|^2 u + V(r) \psi 
\label{reqn2} 
\eeq
In the case of the untrapped condensate, and particularly on a homogeneous background density $\psi_{0}^{2}$, an exact analytical solution of Eq. (\ref{reqn2}) in the form of a rectilinear dark soliton along the $x$-direction reads \cite{luther}, 
\begin{equation}
\psi(x,t)=\psi_{0}\left( \cos \varphi \tanh \zeta +i\sin
\varphi \right)\exp (-i\mu t),  
\label{ds}
\end{equation}
where $\mu \equiv \psi_{0}^{2}$ is the chemical potential, $\varphi$ is the phase shift 
($|\varphi |<\pi /2$) across the dark soliton, $\zeta \equiv \psi_{0}\left( \cos \varphi \right) \left[x-\psi_{0}(\sin \varphi )t\right]$, while the amplitude and velocity of the soliton are given by 
$\cos \varphi $ and $\sin \varphi $ respectively; note that the limit case $\varphi=0$ corresponds to a stationary dark soliton, $u=\psi_{0}\tanh (\psi_{0}x)\exp (-i\mu t)$, often called ``black soliton''. In the same case (i.e., in the absence of the potential),  
the transverse modulational instability of the dark soliton in Eq. (\ref{reqn2}) occurs for perturbation wavenumbers 
\beq
Q < Q_{cr} \equiv  
\left[ 2 \sqrt{\sin^2{\phi}+ \psi_{0}^{-2}\sin{\phi}+
\psi_{0}^{-4}}-\left(2\sin{\phi}+\psi_{0}^{-2} \right)\right]^{1/2}.
\label{qcr}
\eeq
Equation (\ref{qcr}) shows that the instability band may be suppressed if a dark stripe is bent so as to form a ring of length $L<2\pi Q$. This argument led to the concept of ring dark solitons (RDS) \cite{yuri1}, which (under certain conditions) are not subject to the SI. The RDSs were first introduced in the context of optics, where they have been studied both theoretically \cite{yuri2} and experimentally \cite{yuri3}, while, later, they were also predicted to occur in BECs \cite{prlrds}. Importantly, in the latter context, the presence of the magnetic trapping potential is responsible for the formation of robust vortex 
necklaces, when the RDS are subject to the  snaking instability. 

\subsection{Vortex Necklaces formed due to the Snaking Instability}

A ring dark soliton solution of Eq. (\ref{reqn2}) can be sought in the form $\psi=\psi_{TF} \upsilon$ \cite{prlrds}, where 
\beq
\psi_{TF}=\sqrt{\mu -\left(
1/2\right) \Omega ^{2}r^{2}}\exp \left( -i\mu t\right), 
\eeq
is the approximate ground state of the system (Thomas-Fermi cloud) and $\mu$ 
is the chemical potential; also, consistently with the form of the rectilinear dark soliton in Eq. (\ref{ds})), $\upsilon$ is the wave function of the RDS, given by   
\beq
\upsilon (r,t)=\cos \varphi (t)\cdot \tanh
\zeta +i\sin \varphi (t),
\eeq
where $\zeta \equiv \cos \varphi (t)\left[ r-R(t)\right]$, while $\varphi (t)$ and $R(t)$ are the time-dependent phase and radius of the RDS. 
Then, the adiabatic perturbation theory for dark solitons \cite{luther,prlrds} can be used to obtain (for $\mu=1$ and almost black solitons with $\cos \varphi \approx 1$), the following equation of motion for the time-dependent 
ring radius, 
\beq
\frac{d^{2}R}{dt^{2}}=
-\frac{1}{2}\frac{dW(R)}{dR}+\frac{1}{3R}.
\eeq
This equation implies that the RDS moves in a combined effective potential \cite{prlrds} 
$\Pi (R)=(1/2)(\Omega R)^{2}-(1/3)\ln R$. Furthermore, there is a critical radius  
$R_c=\sqrt{2/(3 \Omega)}$, 
for which the RDS in the 
BEC is stationary (unlike its optical counterpart which 
expands indefinitely \cite{yuri1}); this radius 
can also be obtained through solvability conditions
\cite{kaper}. For any $R \neq R_c$, the ring will execute oscillations 
between a minimum and a maximum radius in the potential $\Pi(R)$ \cite{prlrds}, 
with the frequency of small oscillations around the minimum being $\Omega$.

\begin{figure}[th]
\epsfig{figure=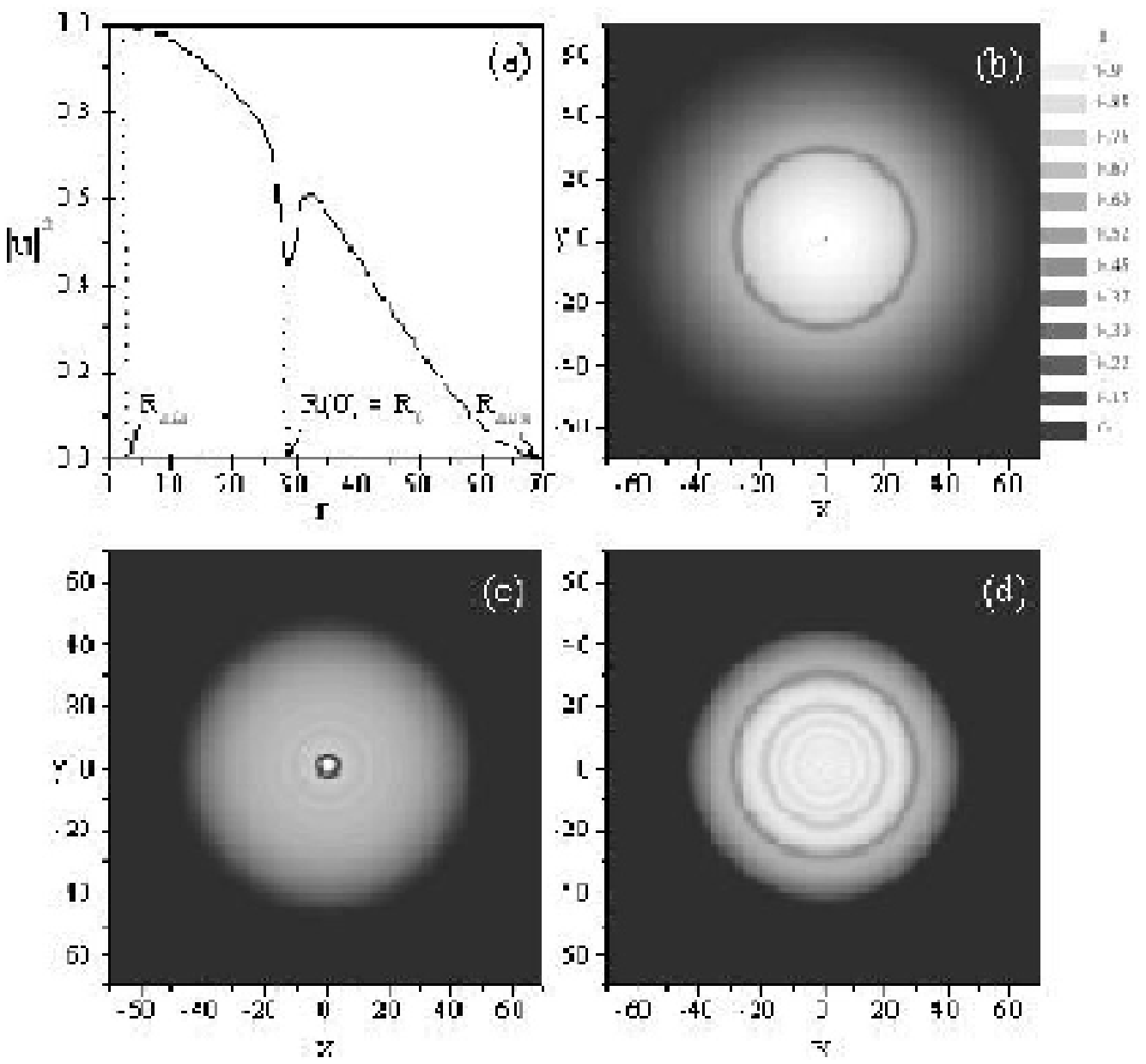,height=2.4in,angle=0}
\epsfig{figure=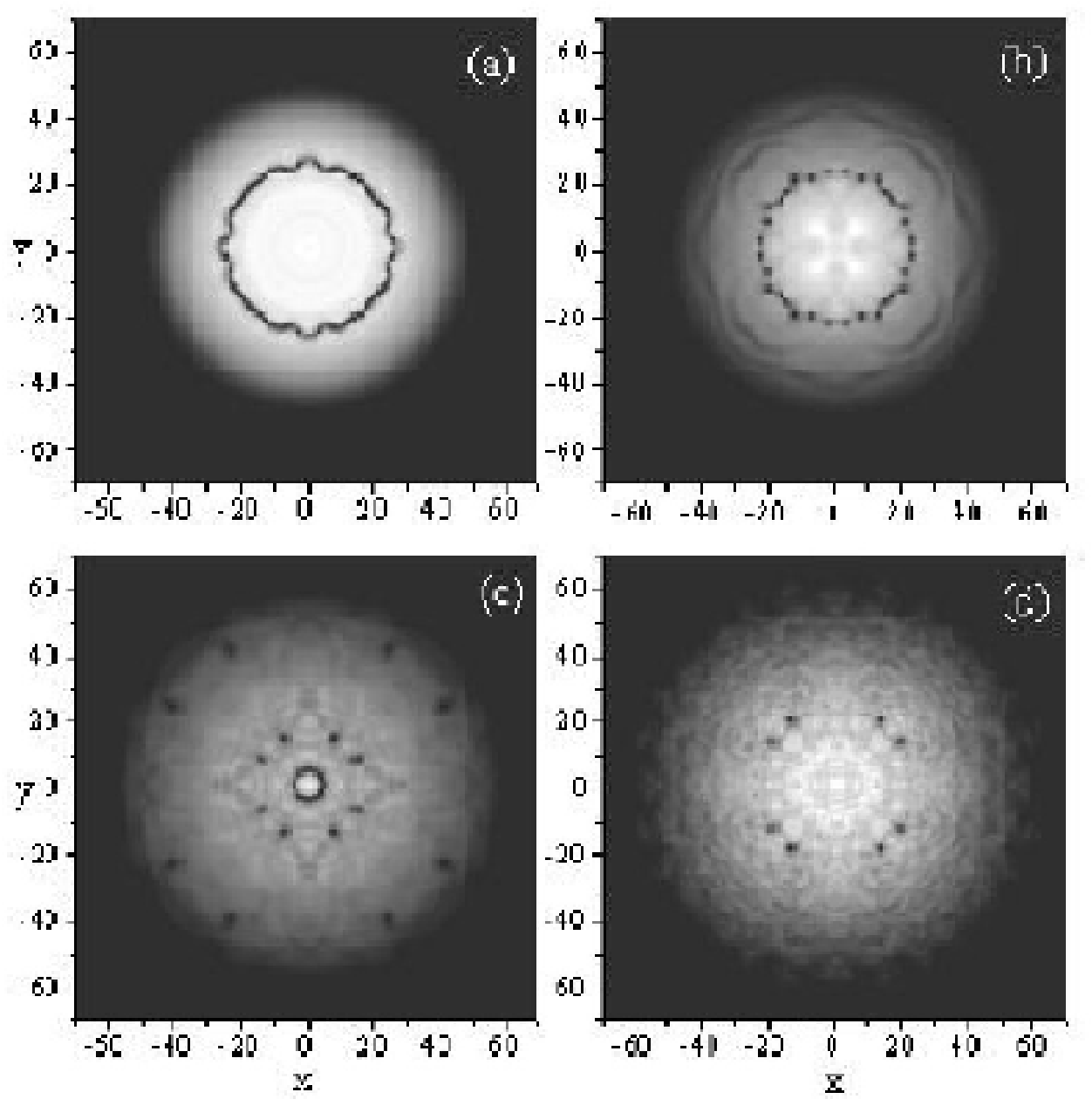,height=2.4in,angle=0}
\vspace*{8pt}
\caption{Left panel: Evolution of a moving (shallow) ring dark soliton.
The top left subplot shows a cross-section of the radial dependence
of the wavefunction (TF+RDS). The top right, bottom left and bottom
right subplots show the evolution of a ring, as it shrinks and subsequently
expands.
Right Panel: Evolution of a deep, originally stationary ring soliton,
breaking up due to the snaking instability into (eventually) 4 vortex
pairs.}
\label{rfig8}
\end{figure}

However, while for shallow rings (initially placed near the BEC center) the dynamics follows the above effective description (cf. left panel of Fig. \ref{rfig8}), for sufficiently deep rings, the RDS dynamics is much more complex. In particular, the rings become unstable to azimuthal perturbations and, thus, subject to the 
SI that results in their breakup. This, in turn, leads to the formation of vortex-antivortex
pairs. How many such pairs will be formed originally is a function of the soliton depth. In \cite{prlrds}, it was found that $8$, $16$, $24$ or $32$ vortices could initially be formed, from the RDS breakup,
but eventually only 8 vortex patterns would be selected by the long-time asymptotics
of the system (cf. right panel of Fig. \ref{rfig8}). The latter were found to dynamically alternate between elaborate x- and cross-shaped patterns \cite{prlrds}.

\subsection{Avoiding the Transverse Instability}


Similarly to the case of the modulational instability in 
quasi-$1$D condensates (cf. section 4.4), in this subsection, we aim to 
illustrate the criterion for suppressing the transverse instability of 
dark solitons in quasi-$2$D BECs (see relevant work in \cite{shlyap} and from that perspective also in \cite{andrea}). Furthermore, we will test this criterion against direct numerical simulations, exposing the possible dynamical scenarios and quantifying their dependence on the trapping parameters. 
%
  
\begin{figure}[th]
\epsfig{figure=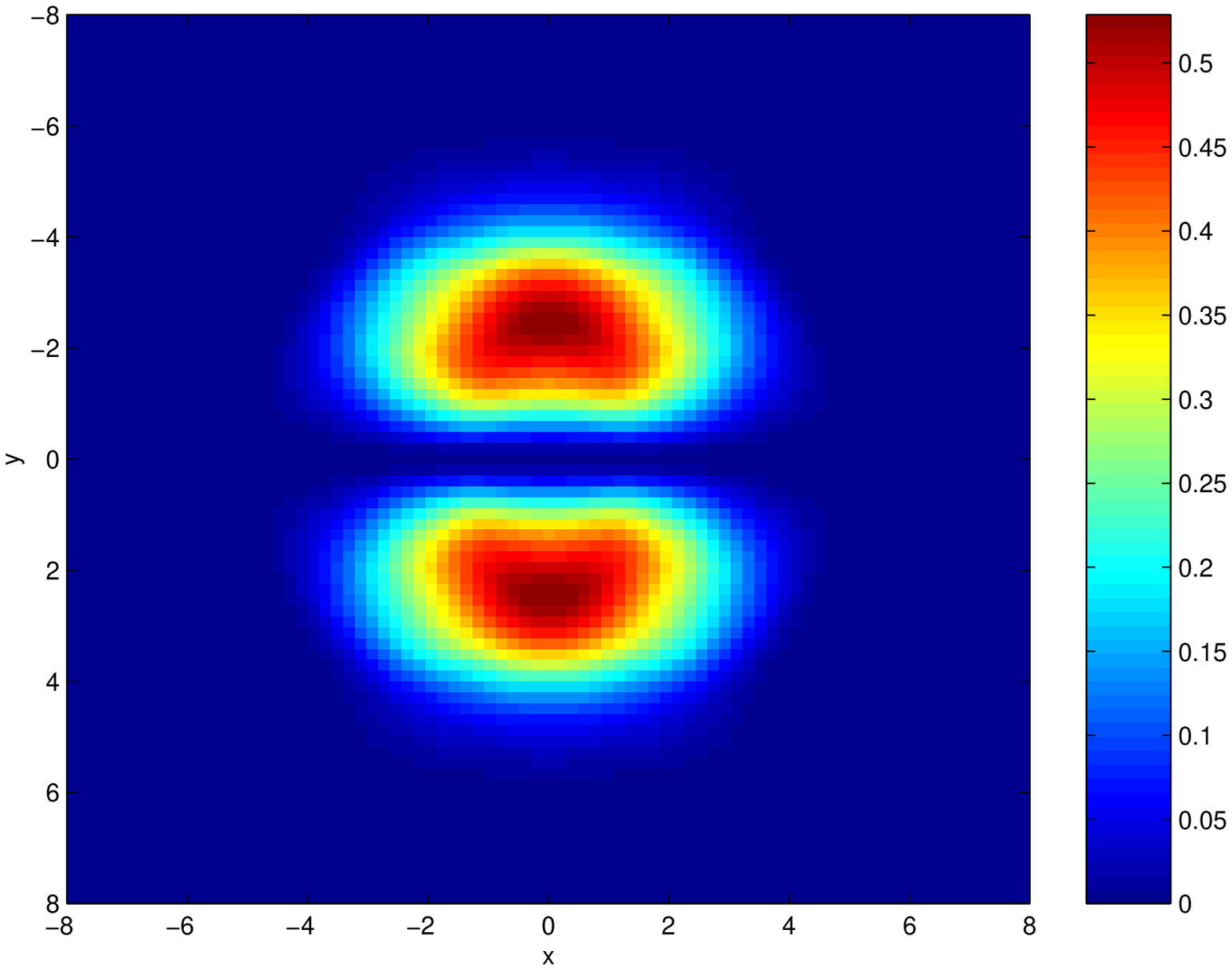,height=1.9in,angle=0}
\epsfig{figure=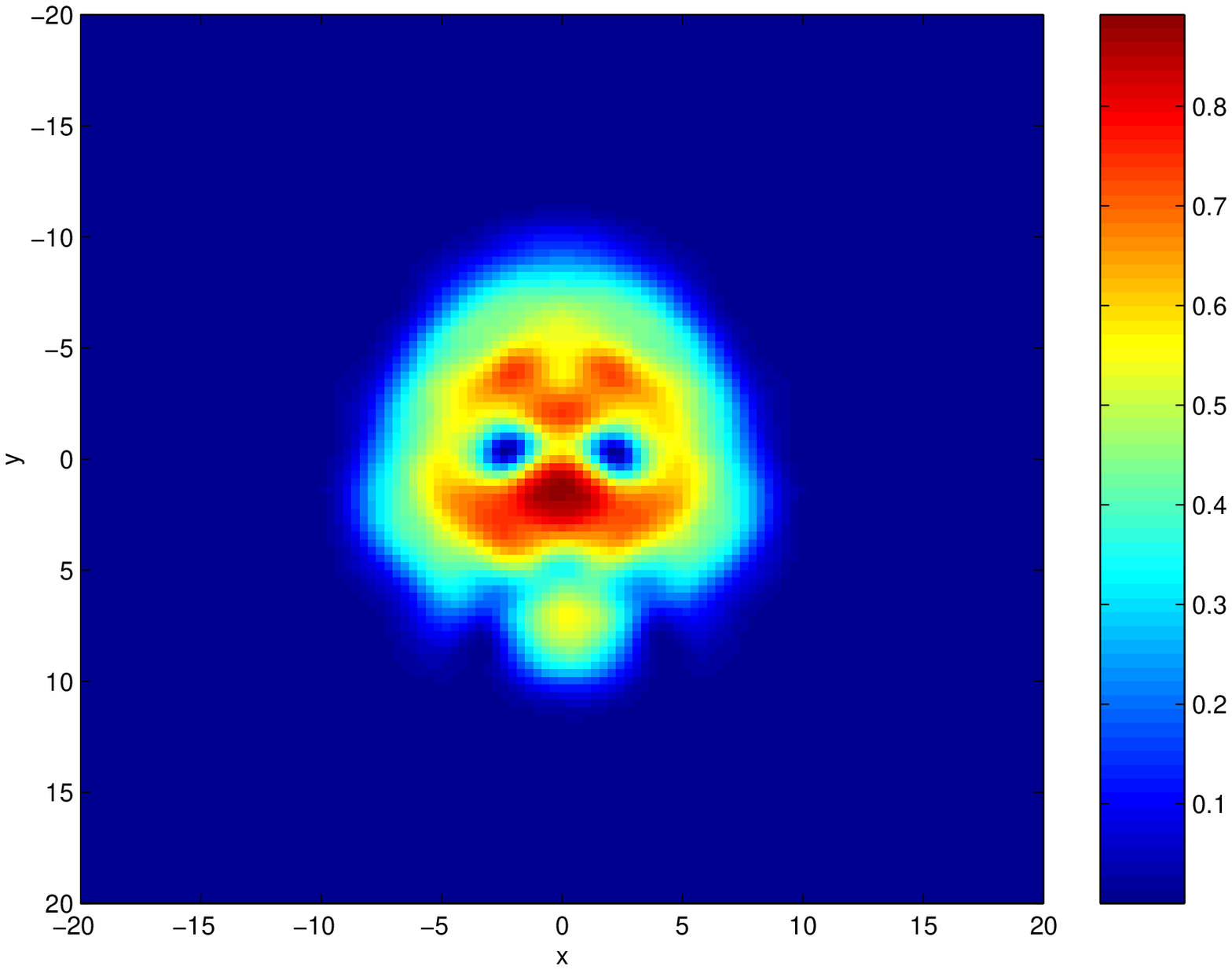,height=1.9in,angle=0}
\vspace*{8pt}
\epsfig{figure=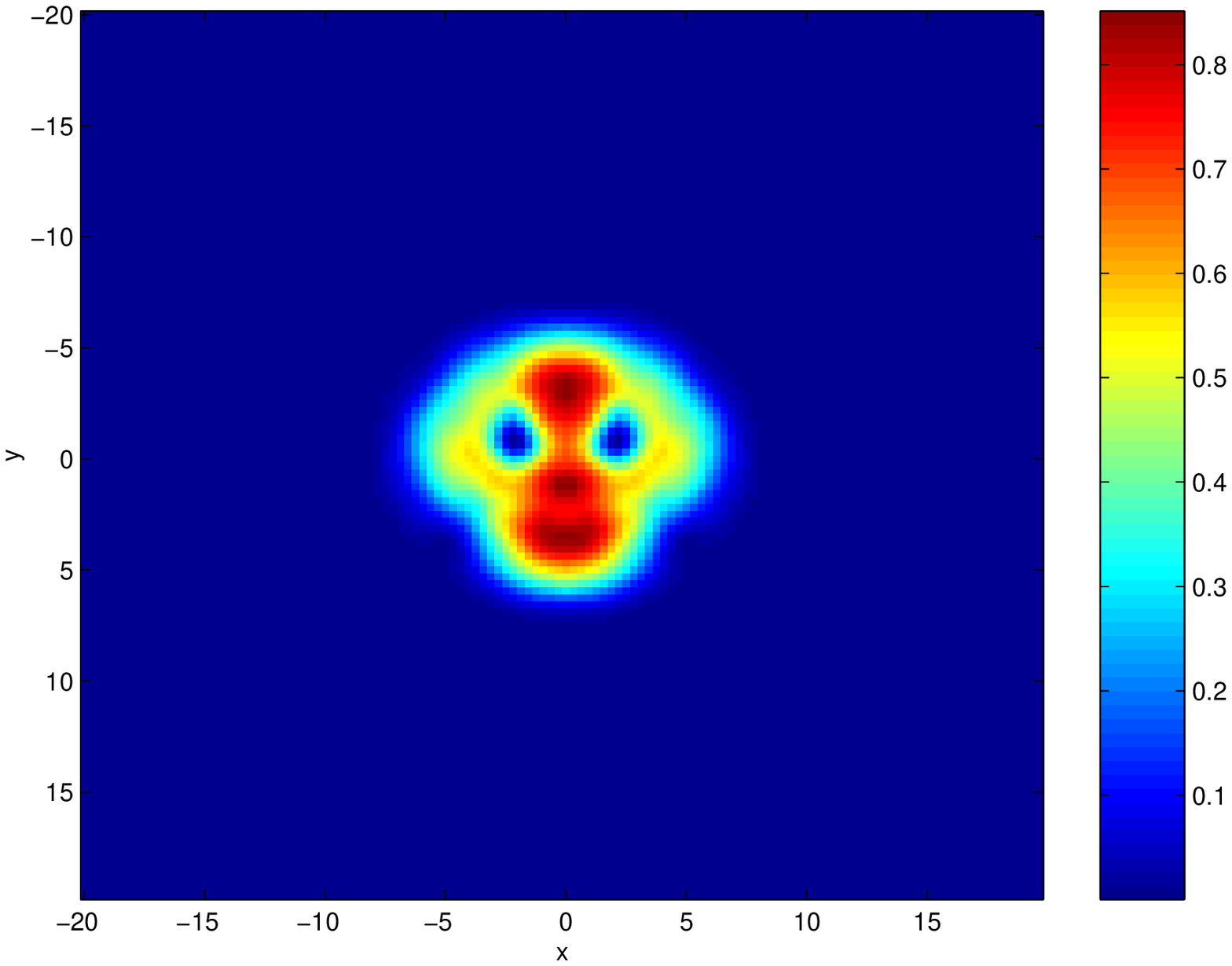,height=1.9in,angle=0}
\epsfig{figure=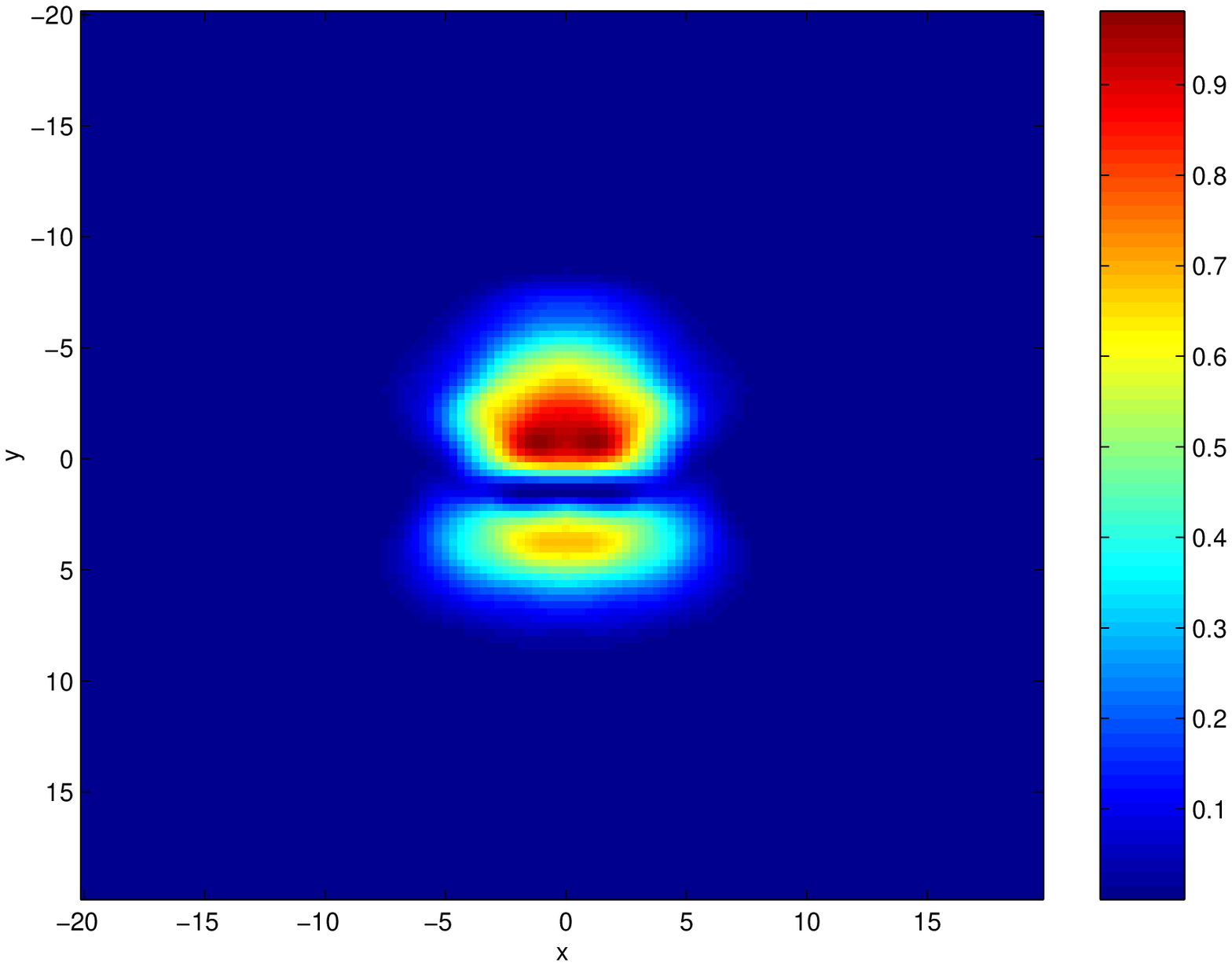,height=1.9in,angle=0}
\vspace*{8pt}
\caption{Top: the panels show the contour plots of the density  
$|\psi|^2$ for $\Omega=0.35$ at $t=1000$ (left) and  
$\Omega=0.15$ at $t=200$ (right). In the first case the  
transverse instability is clearly suppressed, while  
in the second it sets in, giving rise to a formation of a vortex pair. 
Bottom: Snapshots of a vortex-pair evolution in a case  
where snaking instability has set in ($\Omega=0.2$).  
In the left panel ($t=190$) the formed vortex pair is shown,  
while the right panel ($t=210$) shows the recombination  
of the two vortices, resulting in the re-creation of a dark stripe structure.  
The latter is unstable and decays at longer times ($t\approx 400$).}
\label{rfig9}
\end{figure}

To proceed in this direction, we confine ourselves to the case of stationary (black) solitons with $\sin\phi=0$, for which Eq. (\ref{qcr}) leads to $Q_{cr}=\psi_{0}^{-1}$ (recall that $\psi_{0}^{2}$ is the density of an untrapped BEC). On the other hand, in the case of a harmonic trapping potential $V(r)=(1/2)\Omega^2 r^2$, a similar calculation as for the $1$D problem yields the characteristic length scale of the BEC (i.e., the diameter of the TF cloud) as $\lambda_{BEC} \approx 2 \sqrt{2\mu}/\Omega$. Then, the criterion for the suppression of the transverse instability is that the scale of the BEC is shorter than the minimal one for the instability. The corresponding condition reads 
\begin{eqnarray} 
\Omega > \frac{\sqrt{2\mu}}{\pi \psi_{0}}. 
\label{eqn6} 
\end{eqnarray} 
To obtain the minimum value of $\Omega$ we need to know how the chemical potential $\mu$ is connected with $\psi_{0}$. As a first guess, in the absence of the dark soliton, one can assume $\psi_{0}^{2} \approx \mu$ 
(close to the center of the BEC), which yields $\Omega > \sqrt{2}/\pi = 
0.45$. Hence, stronger trapping should ``drown'' the transverse instability 
and preserve dark soliton stripes on top of the Thomas-Fermi cloud  
(i.e., stable ``dipole'' solutions). Note that in terms of real  
physical units, the above mentioned critical value of $\Omega$  
may correspond, e.g., to a weakly interacting $^{87}$Rb pancake  
condensate, containing $\approx 10^{3}$ atoms, confined in a trap  
with $\omega_{r}=2\pi \times 5$Hz and $\omega_{z}=2\pi \times 50$Hz. 
 This condition was numerically tested in \cite{andrea} and was
found to be an  
{\it overestimate} of the critical trapping frequency  
for the transverse instability, of $\Omega_{cr}  
\approx 0.31$.     
It was also found in \cite{andrea} that 
for $0.18 \lesssim \Omega \lesssim 0.31$, while the stripe was dynamically  
unstable, there is not sufficient space  
for the instability-induced vortices to fully develop; as a result,  
after their formation, they subsequently recombine and disappear
(cf. bottom panels of Fig. \ref{rfig9}). On the contrary, 
for $\Omega \lesssim 0.18$, 
the vortices will 
survive in the asymptotic evolution of the system, and, naturally, the weaker  
the trapping the larger the number of ``engulfed'' vortices  
generated due to the stripe breakup (cf. top right panel of Fig. \ref{rfig9}).  
A question that naturally arises in the results 
above concerns the disparity between the critical 
point theoretical estimate for the transverse instability 
and the corresponding numerical finding. 
In \cite{andrea}, this disparity was attributed to the fact  
that the theoretical stability analysis of \cite{luther} is  
performed for the infinite homogeneous medium,  
in the absence of a magnetic trap. In the presence of the trap   
on the one hand, the background is inhomogeneous, 
while, on the other hand,
for tight traps resulting in small condensate sizes, 
the presence of the dark soliton at the BEC center, 
significantly modifies the maximum density. 
Thus, one should expect that the relation $\psi_{0}^{2}=\mu$ 
should be modified as $\delta \psi_{0}^{2}=\mu$, where 
the ``rescaling'' factor $\delta<1$. The numerical simulations  
indicated that embedding the dark soliton in the BEC 
center reduces the maximum density of the TF cloud to 
half of its initial value. 
This suggests that $\delta=1/2$, which, in turn, 
leads to a new minimum value of $\Omega$, 
namely $\Omega=1/\pi\approx 0.318$. 
This modified criterion for the suppression of the 
transverse instability is in a 
very good agreement with the numerically 
obtained  condition of $\Omega>0.31$. 
 
\section{Outlook}

In this short review, we have tried to touch upon a number of aspects of certain (``pattern forming'') dynamical instabilities that Bose-Einstein condensates can sustain. The main message that we have tried 
to convey is that the mean field models of BECs result into nonlinear partial differential equations of the NLS type, about which a wealth of features is inherited from earlier
studies in nonlinear optics and nonlinear wave theory. However, this new exciting world of matter-waves also offers a number of interesting twists (such as the default inclusion of trapping potentials, or the external control via magnetic fields) that can be interwoven and may considerably modify the previously developed body of knowledge.

A particularly exciting feature that we chiefly tried to explore in this novel context of BECs in the present work, is the potential of dynamical transitions that, in turn, lead to the formation of patterns. Such patterns may range from matter-wave bright soliton trains and bound states in quasi-one dimensional systems to vortex compounds and vortex lattices in higher-dimensional settings. The experimental realization of such structures 
\cite{latt1,latt2,latt3,bright1,njp2003b,njp2003} has really motivated and intensified the detailed examination of nonlinear waves in these contexts. Matter-wave bright  
\cite{bright1} and dark \cite{lewenst,prarric} solitons, twisted solitons \cite{njpkevr,louis},
gap solitons \cite{salernoo,ostrovsk}, Feshbach solitons \cite{FRM,dep}, propeller domain-wall solitons \cite{propeller}, shock waves \cite{damsk1,damsk2,damsk3}, vortices 
in the presence \cite{jpbpgk,morch2} 
or the absence \cite{rokshar,feder} of optical lattices, interactions of solitons with sound \cite{prouk} or with other solitons \cite{kivarx,sakaguchi} are only a few examples of structures that have been studied in this context. A useful resource letter containing many relevant references (especially for vortices and various excitation modes of BECs) can be found in \cite{dshall}.

In conclusion, nonlinear waves and patterns have really
found a new paradigm in the context of Bose-Einstein
condensates. Yet, numerous aspects of this general direction
still need to be clarified. Such examples include (among many
others):
\begin{itemlist}
\item aspects of the interaction of the condensate with the gas for
nonzero temperatures \cite{griffin,law,pablo};
\item features of the collapse type phenomena that can occur in
higher dimensional settings (where experimental results still challenge
the theoretical interpretations \cite{feshb1});
\item multi-component condensates where numerous novel structures
still await to be discovered/observed \cite{myatt,bernard,epjd},
\item as well as fundamental instabilities such as the Landau
instability, that have sparked a long standing controversy
that still waits to be undisputably resolved (see e.g., the
recent discusion of \cite{wuniunjp} and references therein).
\end{itemlist}
This exciting journey has just begun $\dots$

\vspace{5mm}

{\bf Acknowledgements}

It is a pleasure and an honor to acknowledge the invaluable contribution of 
our collaborators in many of the topics discussed here: Egor Alfimov, 
Alan Bishop, Ricardo Carretero-Gonz\'alez, Fotis Diakonos, Todd Kapitula, 
Yannis Kevrekidis, Yuri Kivshar, Volodya Konotop, Boris Malomed, 
Hector Nistazakis, Dmitry Pelinovsky, Mason Porter,
Nick Proukakis, Mario Salerno, 
Peter Schmelcher, Augusto Smerzi and Andrea Trombettoni are greatly thanked. 
We would also like to acknowledge numerous 
useful discussions with Peter Engels, David Hall and Alexandru Nicolin.
We are grateful to the Florence group for kindly permitting us to use
their experimental results in Fig. \ref{rfig6}.
Last but not least, we also thank our students Zoi Rapti (UMass) and George Theocharis (Athens) for their keen interest and endless efforts. PGK is grateful to the Eppley Foundation for Research, the NSF-DMS-0204585 and the NSF-CAREER program for financial support.

{\bf Note Added in Proof}

After the submission of this paper, more detailed experimental results
illustrating the appearance of the modulational instability in a 1D 
moving optical lattice, in good agreement with the Gross-Pitaevskii
theory, were presented in \cite{flornew}.

\end{document}